\documentclass[conference]{IEEEtran}
\IEEEoverridecommandlockouts
\usepackage{lscape}
\usepackage{cite}
\usepackage{amsmath,amssymb,amsfonts}
\usepackage{siunitx}
\usepackage[linesnumbered,ruled,vlined]{algorithm2e}
\usepackage{graphicx}
\usepackage{pgfplots}
\usepackage{svg}
\usepackage{textcomp}
\usepackage{ulem}
\usepackage{xcolor}
\usepackage{kantlipsum}
\usepackage{verbatim}
\usepackage{todonotes}
\usepackage[usestackEOL]{stackengine}
\usepackage{flushend}
\usepackage{breqn}

\usepackage{float}

\usepackage{caption}
\usepackage{subcaption}
\usepackage[para]{threeparttable}
\usepackage{microtype}

\def\BibTeX{{\rm B\kern-.05em{\sc i\kern-.025em b}\kern-.08em
    T\kern-.1667em\lower.7ex\hbox{E}\kern-.125emX}}
\DeclareSIUnit\op{OP}
\DeclareSIUnit\ops{OPS}
\DeclareSIUnit\mac{MAC}
\DeclareSIUnit\macs{MACS}
\DeclareSIUnit\cycle{cycle}
\definecolor{customblue}{RGB}{118,159,202}
\definecolor{customorange}{RGB}{255,247,176}
\definecolor{customred}{RGB}{255,108,101}
\pgfplotsset{compat=newest}
\pgfplotsset{
	tick label style = {font = {\fontsize{9 pt}{12 pt}\selectfont}},
	label style = {font = {\fontsize{9 pt}{12 pt}\selectfont}},
	legend style = {font = {\fontsize{9 pt}{12 pt}\selectfont}},
	nodes near coords style = {font = {\fontsize{6 pt}{12 pt}\selectfont}},
}
\SetCommentSty{mycommfont}

\SetKwInput{KwInput}{Input}                
\SetKwInput{KwOutput}{Output} 

\begin{document}

\title{Hardware Accelerator and Neural Network Co-Optimization for Ultra-Low-Power Audio Processing Devices\\
\thanks{This work has been partly funded by the EU and the German Federal Ministry of Education and Research (BMBF) in the projects OCEAN12 (reference number: 16ESE0270).}
}

\author{\IEEEauthorblockN{Gerum Christoph\textsuperscript{*}}
\IEEEauthorblockA{\textit{Department of Computer Science} \\
\textit{ University of Tübingen}\\
christoph.gerum@uni-tuebingen.de}
\and
\IEEEauthorblockN{Frischknecht Adrian\textsuperscript{*}}
\IEEEauthorblockA{\textit{Department of Computer Science} \\
\textit{ University of Tübingen}\\
adrian.frischknecht@uni-tuebingen.de}
\and
\IEEEauthorblockN{Hald Tobias}
\IEEEauthorblockA{\textit{Department of Computer Science} \\
\textit{ University of Tübingen}\\
tobias.hald@student.uni-tuebingen.de}
\and
\IEEEauthorblockN{Palomero Bernardo Paul}
\IEEEauthorblockA{\textit{Department of Computer Science} \\
\textit{ University of Tübingen}\\
paul.palomero-bernardo@uni-tuebingen.de}
\and
\IEEEauthorblockN{Lübeck Konstantin}
\IEEEauthorblockA{\textit{Department of Computer Science} \\
\textit{ University of Tübingen}\\
konstantin.luebeck@uni-tuebingen.de}
\and
\IEEEauthorblockN{Bringmann Oliver}
\IEEEauthorblockA{\textit{Department of Computer Science} \\
\textit{ University of Tübingen}\\
oliver.bringmann@uni-tuebingen.de}
}
\maketitle
\begingroup\renewcommand\thefootnote{*}
\footnotetext{These authors contributed equally to this work.}
\endgroup

\begin{abstract}
The increasing spread of artificial neural networks does not stop at ultralow-power edge devices. However, these very often have high computational demand and require specialized hardware accelerators to ensure the design meets power and performance constraints.
The manual optimization of neural networks along with the corresponding hardware accelerators can be very challenging. This paper presents HANNAH (Hardware Accelerator and Neural Network seArcH), a framework for automated and combined hardware/software co-design of deep neural networks and hardware accelerators for resource and power-constrained edge devices. The optimization approach uses an evolution-based search algorithm, a neural network template technique and analytical KPI models for the configurable UltraTrail hardware accelerator template in order to find an optimized neural network and accelerator configuration.
We demonstrate that HANNAH can find suitable neural networks with minimized power consumption and high accuracy for different audio classification tasks such as single-class wake word detection, multi-class keyword detection and voice activity detection,
which are superior to the related work.
\end{abstract}

\begin{IEEEkeywords}
Machine Learning, Neural Networks, AutoML, Neural Architecture Search
\end{IEEEkeywords}
\bstctlcite{IEEE:BSTcontrol}
\section{Introduction}\label{sec:introduction}

Artificial intelligence is increasingly spreading into the domain of always-on ultra-low
power connected devices like fitness trackers, smart IoT sensors, hearing aids and smart speakers. 
The limited power budget on these devices and the high computational demand 
often mandates the use of specialized ultra-low power hardware accelerators, 
specialized to a specific application or application domain. 
Hardware design, neural network training and optimized deployment often
require manual optimization by the system designers, who need to deal with manifold often counter-directed issues.
In this work, we propose HANNAH (Hardware Accelerator and Neural Network seArcH) 
to automatically co-optimize neural network architectures and a corresponding 
neural network accelerator. 

\begin{figure}[t]
	\centering
	\includegraphics[width=\linewidth]{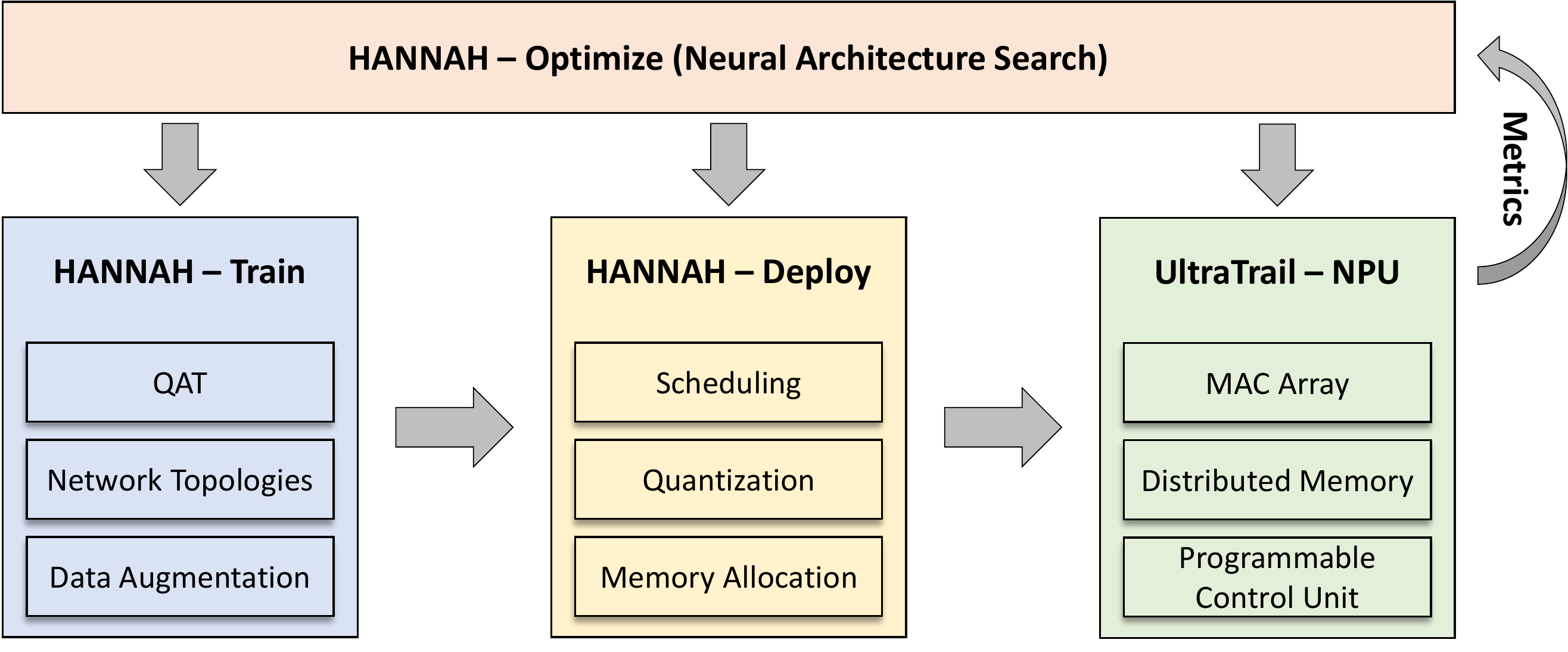}
	\caption{Overview of the HANNAH framework.}
	\label{fig:hannah_overview}
\end{figure}

The HANNAH design flow is shown in Figure~\ref{fig:hannah_overview}. 
Neural Networks are instantiated and trained in the training  
component employing quantization aware training 
and dataset augmentation. Trained neural networks are then handed 
over to the deployment component (Sec.~\ref{sec:deployment}) for 
target code generation. Here, the neural network is quantized to a 
low word width representation the neural network operations are scheduled 
and on-device memory is allocated for the neural network. Along with 
the target network architecture, a specialized hardware accelerator for the 
neural network processing is instantiated from a configurable Verilog template. 
Hardware dependent performance metrics like power consumption and chip area 
are then either generated by running the neural network on a 
gate-level  simulation or estimated using an analytical performance model.
The evolution-based search strategy (Sec.~\ref{sec:optimization}) implemented in 
the HANNAH optimizer is then used to incrementally search the neural 
network and hardware accelerator codesign space. 

The main contributions of this paper are:

\begin{enumerate}
\item We present an end-to-end design flow from neural network descriptions down to synthesis 
  and gate-level power estimation. 
\item We propose a model-guided hardware and neural network co-optimization and architecture exploration flow for ultra-low power devices.
\item The combined flow reaches state-of-the-art results on a variety audio detection tasks. 
\end{enumerate}

\section{Related Work}\label{sec:related}
In recent years, neural architecture search (NAS) had an enormous success \cite{pham2018efficient, zoph2018learning, liu2017hierarchical} 
in automating the process of designing new neural network architectures. 
The early work 
on neural architecture search did not take hardware characteristics into account. More 
recent works (Hardware-Aware NAS) allow to take the execution latency of a specific 
target hardware into account and allow to optimize the neural networks for latency 
as well as accuracy.
There are several approaches applying hardware/software co-design to hardware accelerators 
for neural networks. In early works this involves manually designing specialized 
neural network operations and a hardware architecture \cite{yang2019synetgy}. 
In \cite{abdelfattah2020best} and \cite{jiang2020hardware} reinforcement learning based 
NAS is extended to include search for an accelerator configuration on FPGAs and 
optimize it for latency and area. 
Zhou et al. \cite{zhou2021rethinking} for the first time combine differentiable 
neural architecture search and the search for a neural architecture configuration.
All of these approaches are not directly applicable to TinyML, as their search spaces for
neural networks and hardware accelerators make it impossible to meet power budgets in the order
of \SI{<10}{\micro\watt}.

NAS for TinyML mostly focuses on searching neural networks
for microcontrollers. The search methods in \cite{liberis2021munas, fedorov2019sparse} 
use genetic algorithms and Bayesian optimization to optimize neural networks to fit the 
constrained compute and memory requirements of these devices. 
In \cite{lin2020mcunet} a weight sharing approach is used to train a super network 
containing multiple neural networks at once and a genetic algorithm, is 
used to search for the final network topology. 
A first approach to apply 
differentiable neural architecture search to TinyML is presented in 
\cite{banbury2021micronets} it reaches state-of-the-art results on the tinyMLPerf 
benchmarks, but requires relatively big microcontrollers (\raisebox{-0.8ex}{\~{}}500kB of memory) to execute 
the found networks. These search methods would generally be applicable to our neural network search problem,
but they do not contain a co-optimization strategy for also optimizing the hardware architecture. 

For ultra-low power audio processing hardware accelerators or other edge devices with our intended target power there 
are currently no hardware accelerators. So our main state of the art comprises of manually designed hardware accelerators for 
a specific neural network architecture. Manual optimization of ultra-low power hardware has been used for all of the use-cases in the 
experimental evaluation.  
Recent examples include keyword spotting (KWS) \cite{bernardo2020ultratrail, giraldo2018laika,giraldo_vocell_2020,liu_ultra-low_2019,price_low-power_2018}, 
wake word detection (WWD) \cite{wang2020always, guo20195} and voice activity detection (VAD)
\cite{price_low-power_2018, yang2019design,liu2020background}. These approaches show 
impressive results, but all require a work intensive and error-prone manual design process.


\section{NPU Architecture and Analytical Modeling}\label{sec:deployment}
\subsection{NPU architecture} \label{sec:npu_architecture}
\begin{figure}[!t]
	\centering
	\includegraphics[width=\linewidth]{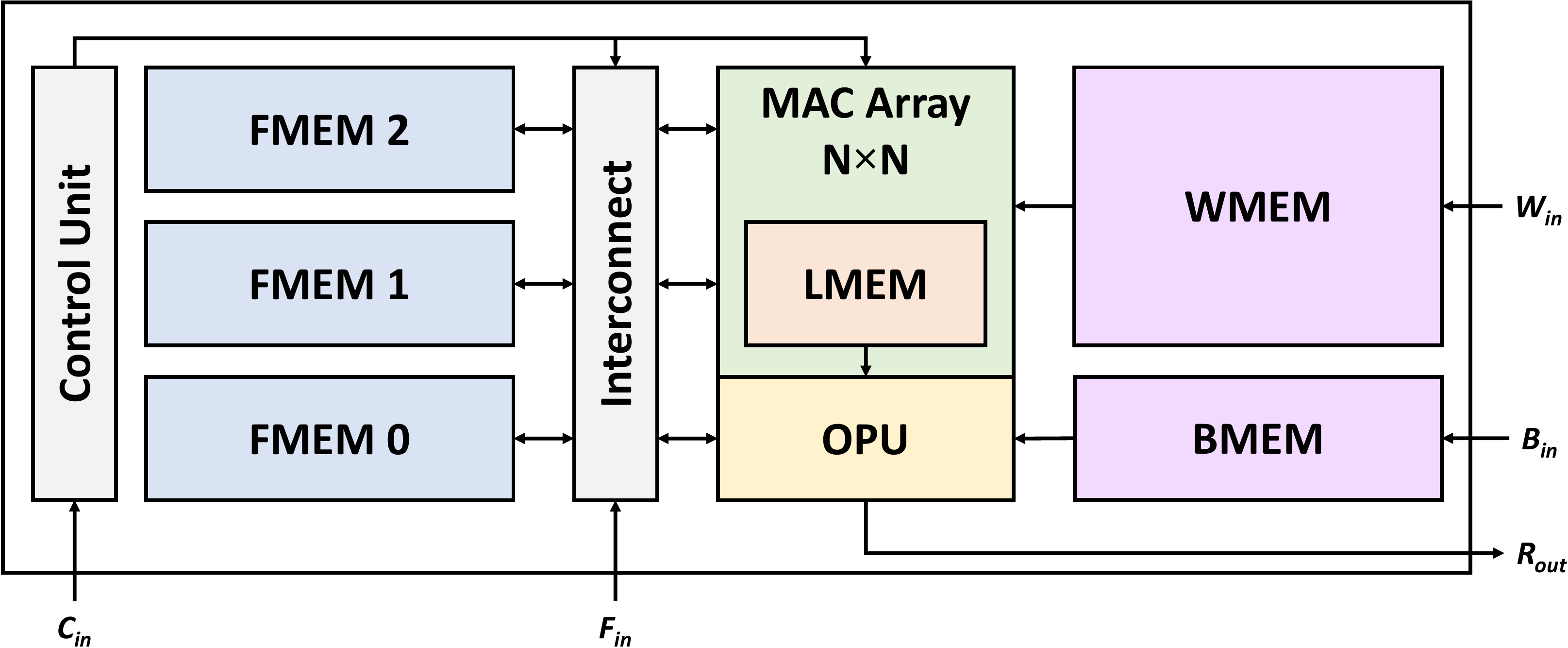}
	\caption{Overview of the UltraTrail architecture.}
	\label{fig:ut_overview}
\end{figure}

The NPU architecture used in this work is based on UltraTrail \cite{bernardo2020ultratrail}, a configurable accelerator designed for TC-ResNet.
Fig. \ref{fig:ut_overview} shows the basic UltraTrail architecture. The accelerator uses distributed memories to store the features (FMEM0-2), parameters (W/BMEM) and local results (LMEM). Features, parameters and internal results use a fixed-point representation.
An array of multiply and accumulate (MAC) units calculates the convolutional and fully-connect layers.
A separate output processing unit (OPU) handles post-processing operations like bias addition, application of a ReLU activation function, average pooling.
A configuration register stores the structure of a trained neural network layer by layer. Each layer configuration contains, among others, information about the input feature length, input and output feature channels, kernel size, and stride, as well as the use of padding, bias addition, ReLU activation, or average pooling.

The architecture of UltraTrail has been parameterized to suit the executed neural networks as best as possible. At design time, size and number of supported layers of the configuration register, the word width for the features, parameters, and internal results, the size of the memories, the dimension of the quadratic MAC array, and the number of supported classification classes can be modified. During runtime, the programmable configuration register allows the execution of different neural networks.

\subsection{Deployment}

To execute a trained neural network with UltraTrail, a schedule, a corresponding configuration, and a sorted binary representation of the quantized weights and bias values have to be created. These steps are part of the automated deployment flow.

The deployment backend converts the PyTorch Lightning model from HANNAH to ONNX and its graph model representation. A graph tracer iterates over the new representation and determines the schedule and the layer-wise configuration which is loaded into the configuration register of UltraTrail. The tracer changes the node order in the obtained graph such that residual connections are processed before the main branch to ensure a correct schedule for the NPU as described in \cite{bernardo2020ultratrail}. The graph tracer also searches for exit branches for early exiting to handle correct configuration and treatment when the exit is not taken.

The backend extracts weights and biases from the model. These parameters are padded and reordered to fit the expected data layout before they get quantized to the defined fixed-point format. The same happens to example input data provided by HANNAH which is used as simulation input.

The word widths of the internal LMEM memory are calculated by the largest fixed-point format times two plus the logarithm of the maximum input channels to avoid an overflow.

After the schedule and number of weights and biases are fixed, the deployment backend generates the corresponding 22FDX memory macros. The backend determines the size and word width of the weight and bias memory from the fixed numbers. The schedule gets analyzed to determine the size of each FMEM so that the feature maps fit perfectly for the trained neural network. All memory sizes can be adjusted manually to support other neural networks with more parameters or larger feature maps. The LMEM size is determined by the largest supported output feature map. All memory sizes and word widths are set to the next possible memory macro configuration.

Given the generated configuration, weights and biases, memory macros, example input data, and selected hardware parameters, a simulation, synthesis, and power estimation are run automatically if desired. The simulation results are fed back to HANNAH and compared with the reference output to validate the functional correctness of the accelerator.



\subsection{NPU Models}\label{sec:npu_models}

The simulation and synthesis of UltraTrail are time-consuming and therefore not feasible for hundreds to thousands of possible configurations. 
To get an accurate yet fast evaluation of UltraTrail,  performance, area, and power consumption are estimated using analytical models.

\normalem
\begin{algorithm}[t]
	\For{$\mathtt{k=0:\lceil K/N\rceil}$}{
	  \For{$\mathtt{c=0:\lceil C/N \rceil}$}{
		\For{$\mathtt{f= 0:F}$}{
		  \For{$\mathtt{x=0:X}$}{
        	$\mathtt{i_{idx} \gets x\cdot S-\lfloor F/2 \rfloor+f}$ \\
            \If{$\mathtt{i_{idx} \ge 0}$ and  $\mathtt{i_{idx} < I}$}{
				$\mathtt{l^{(N)}[x]\mathrel{+}= i^{(N)}[c][i_{idx}] \cdot w^{(N \times N)}[k][c][f]}$ \\
			}
		  }  
		}
	  }
	  \For{$\mathtt{x=0:X}$}{
		$\mathtt{o^{(N)}[k][x] = opu(l^{(N)}[x])}$
	  }
	}
	\caption{Pseudocode of the NPU accelerator when configured with an $N\times N$-MAC array with $K$ output channels, $C$ input channels, 
    $F$ filter size, $S$ stride, $I$ input length, $X=\lfloor (I-1)/S\rfloor+1$ output length, and full padding.}
	\label{alg:ultratrail}
\end{algorithm}
\ULforem

For latency estimation we adopt the cycle-accurate analytical model presented in \cite{bernardo2020ultratrail} to configurable array sizes. 
The pseudocode shown in Algorithm~\ref{alg:ultratrail} visualizes the NPU operations and memory 
accesses. The accelerator iterates over $C$ input and $K$ output channels tiled by the MAC array size $N$. One $N \times N$ patch of the weight array is 
fetched from the weight memory per iteration of the next loop. In the innermost loop $N$ input channels are fetched from one of the FMEMs and the current 
convolution outputs are accumulated in the LMEM using a spatially unrolled matrix-vector multiplication on the MAC array. To avoid misaligned memory accesses 
and accelerate the computation, padding is implemented by skipping the corresponding loop iterations instead of actual zero padding of the feature maps.  
In the last loop, the OPU fetches $N$ output channels from the LMEM and the $N$ results are stored in a planned FMEM.
The latency of the accelerator without skipping of padded values can be easily estimated using the following equation, by just
counting the number of loop iterations. 

\begin{equation}
l =  1 + \left\lceil \frac{C}{N} \right\rceil \cdot \left\lceil \frac{K}{N} \right\rceil \cdot F \cdot X
\end{equation}

Where the output length $X$ can be derived from input length $I$ using $X=\lfloor (I-1)/S\rfloor+1$.
The crucial part of the performance model is to accurately calculate the number of skipped loops in Line 7 of Algorithm~\ref{alg:ultratrail}. 
For the case $i_{\mathtt{idx}} \ge 0$ skipping happens in the first $a_{\mathtt{p,b}} = \left\lfloor \frac{\left\lfloor F/2  \right\rfloor - 1}{S} + 1 \right\rfloor$ 
iterations over $x$, the number of skipped executions is $\left\lceil \frac{F}{2} \right\rceil$ at the start of the loop,
and decreases by $S \cdot i$ for each iteration over $X$.
For the case $i_{\mathtt{idx}} < I$ the analytical model is similar but special care must be taken if the input length $I$ is not divisible by the stride.
In the last iteration over $x$, $i_{\mathtt{idx}}$ takes the value $I_{\texttt{max}} = (X-1)\cdot S-\lfloor F/2 \rfloor+F-1$ leading to an effective padding
of $C_{w,e} = I_{\texttt{max}} + 1 -\left\lfloor F/2  \right\rfloor - I$. Loop skipping then happens for the last 
$a_{\mathtt{p,e}} = \left\lfloor \frac{\left\lfloor C_{w,e}  \right\rfloor - 1}{S} + 1 \right\rfloor$ iterations. The number of skipped iterations is
$C_{w,e}$ at the end of the loop and also decreases by $s\cdot i$ at each loop iteration. This leads to an estimation of the total number of skipped MAC array operations 
at the beginning $\#MAC_{not,b}$ and end $\#MAC_{not,e}$ of the loop as:

\begin{equation} 
	\#MAC_{not,b} = \sum^{a_{p,b}-1}_{i=0} \left\lfloor \frac{F}{2} \right\rfloor - s\cdot i
\end{equation}

\begin{equation}
	\#MAC_{not,e} = \sum^{a_{p,e}-1}_{i=0} \left\lfloor C_{w,e} \right\rfloor - s\cdot i
\end{equation}

And a total per layer latency of: 

\begin{equation}
	l =  1 + \left\lceil \frac{C}{N} \right\rceil \cdot \left\lceil \frac{K}{N} \right\rceil \cdot F \cdot X - \#MAC_{not,e} - \#MAC_{not,b}
\end{equation}

The power model has two parts for SRAM power estimation and for other NPU components. For every layer and memory, the model calculates 
the number of read, write, and nop (idle) operations. LMEM and input feature memory (IMEM) are accessed at each cycle of the operation except for a single setup cycle.
The  read cycles $r_{\text{imem}}$ and $r_{\text{lmem}}$ and write cycles $w_{\text{lmem}}$ are the same as the layer latency $l-1$. As the number of memory accesses to the 
weights remain stationary during the innermost loop of Algorithm~\ref{alg:ultratrail}, the weight memory is accessed only  
$\left\lceil \frac{C}{N} \right\rceil \cdot \left\lceil \frac{K}{N} \right\rceil\cdot F$ times ($r_{\text{wmem}}$). 
Memory access to the output feature memory ($w_{omem}$), the bias memory ($r_{bmem}$), and in the case of residual blocks the partial sum feature memory ($r_{pmem}$)
is given as $\left\lceil K/N \right\rceil\cdot X$. Idle times for the memories are then calculated using the layer latencies and access times for each 
memory: $i_{m} = l - r_m - w_m$

Furthermore, the combinational MAC array alone causes many relevant glitches related to SRAM. Therefore, the glitching LMEM data input leads to a non-negligible power increase.
Based on some gate-level simulations for different networks and array sizes, a linear regression for the number of glitches ($g_{lmem}$) is used.

Finally, if the network latency on the accelerator $L$ is below the period $P$ for real time operation, the memories are switched to low power modes with 
reduced leakage for the remainder of the period. The final power consumption of the memories is then estimated using: 


\begin{flalign}
 P_{mem} & = \frac{L}{P} \cdot P^{(glitch)}_{lmem} \cdot g_{lmem} &\\ \nonumber
		 & + \sum_{m} \frac{L}{P}(P^{(read)}_m \cdot r_m + P^{(write)}_m \cdot w_m  + P^{(idle)}_m \cdot i_m)&\\ \nonumber
		 & + \frac{L}{P} \cdot P^{(static)}_m + \left(1-\frac{L}{P}\right) \cdot P^{(lp)}_m \nonumber
\end{flalign}

The dynamic power $P^{(read)}_m$, $ P^{(write)}_m$, $P^{(idle)}_m$, $P^{(glitch)}_{lmem}$ and static power in running-  $P^{(static)}$ and low power-mode $P^{(lp)}$ are 
extracted from the memory compiler for each memory macro usable during the co-optimization and stored in a database.  

Again, the model assumes a constant value for the control unit as the power consumption is mainly independent of the configuration and executed neural network.
The power consumption of the MAC array and OPU is approximated using a linear regression on MAC array size and word width of the MAC array calibrated using gate-level simulations. 
Note, that the dynamic power for non-memory modules is weighted depending on the runtime per inference.

The analytical area model comprises an exact SRAM area calculation and an estimation for the other NPU components to estimate the cell area of the NPU after synthesis. The model looks up the SRAM area in a small database containing all necessary memory macros used during NAS. The SRAM area estimation is by far the most important as it is responsible for about \SI{90}{\percent} of the total synthesis cell area. The control unit is mainly independent of the configuration and its area is modelled as constant. The MAC array is estimated by a linear growth of the word width and quadratic growth of the MAC array dimension based on a minimum MAC unit area. The OPU model is like the MAC array model however grows linear with MAC array size.

\section{Neural Network / Hardware Co-Optimization}\label{sec:optimization}

\begin{figure}[!t]
	\centering
	\begin{subfigure}[t]{.48\columnwidth}
        \vskip 0pt
		\includegraphics[width=\columnwidth]{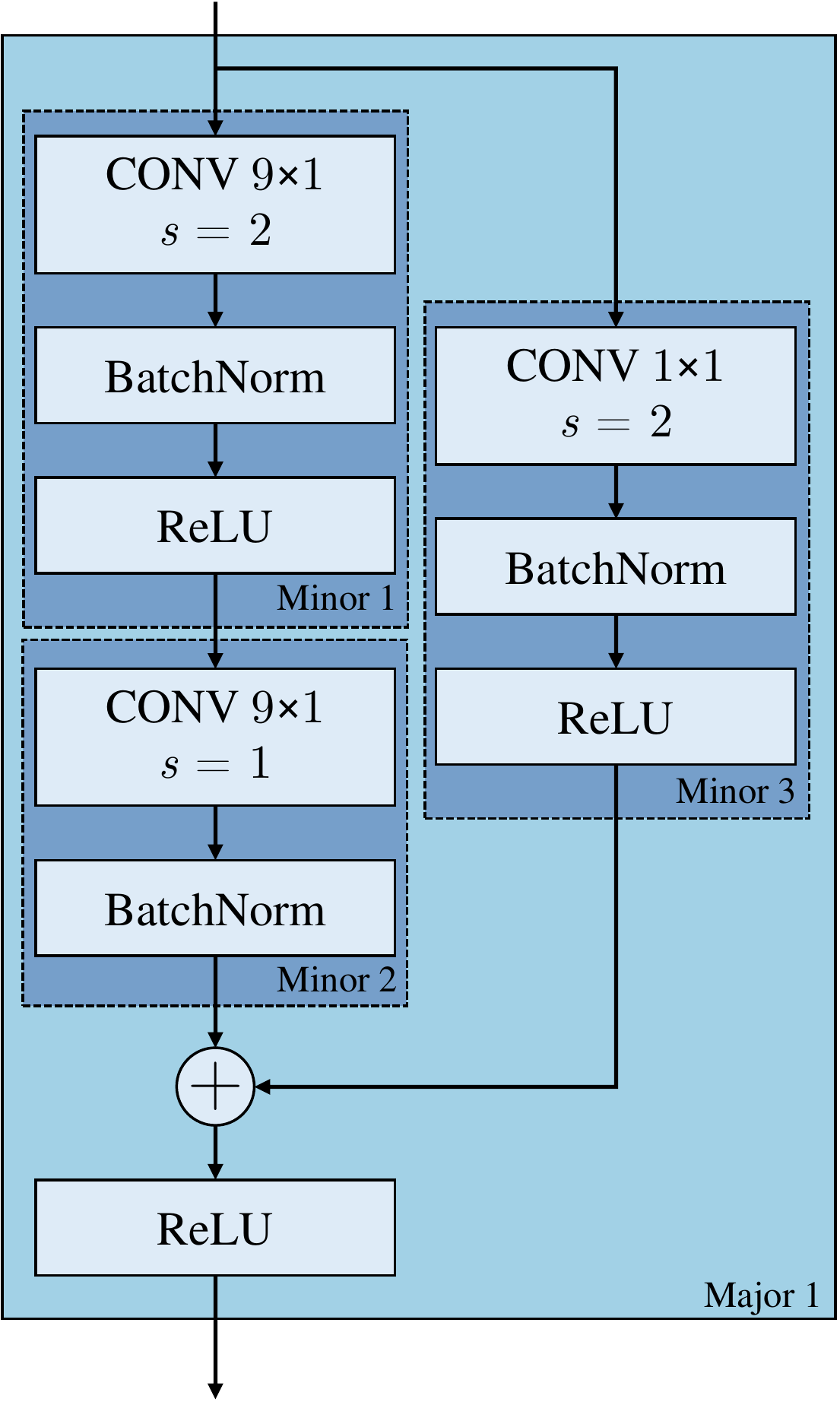}
		\caption{}%
		\label{fig:graph}%
	\end{subfigure}
    \hfill
	\begin{subfigure}[t]{.48\columnwidth}
        \vskip 0pt
		\includegraphics[width=\columnwidth]{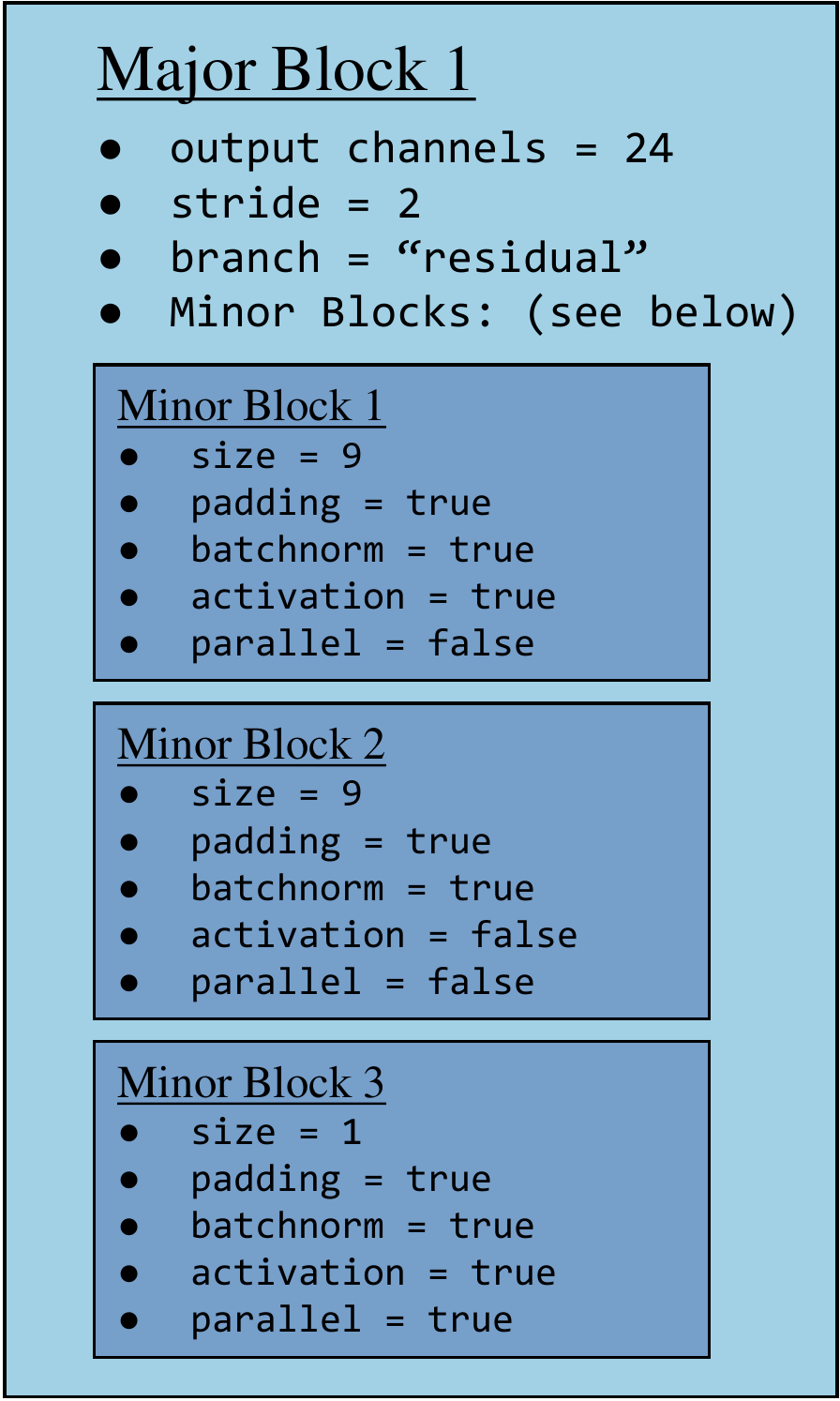}
		\caption{}%
		\label{fig:uml}%
	\end{subfigure}
    \caption{A subgraph of the search space build from blocks and layers (a) and the corresponding abstract configuration (b).}
	\label{fig:majorminor}
\end{figure}

The co-optimization uses a block based search space for neural network architectures. As shown in Figure \ref{fig:graph} each block is either a residual block with a main branch of a configurable number of convolutional layers on the trunk and a skip connection or a simple feed forward CNN block leaving out the skip connection. We additionally search over layer, block and network level hyperparameters like filter size, stride and convolution sizes.
The hardware accelerator search space consists of the MAC array dimensions, the memory sizes and multiplier word widths. The mac array size
is optimized using the search algorithm while the other metrics are derived from the neural network parameters, during neural network deployment.
A full overview of the parameters is shown in Table \ref{tab:search}.

\begin{table}
	\centering
	\begin{threeparttable}
		\caption{Neural Architecture Search Space}
		\label{tab:search}
		\begin{tabular}{c|c|c}
			\hline
			Level & Option & Choices \\
			\hline 
			Network & Word Width Features & 4,6,8  \\
			Network & Word Width Weights & 2,4,6,8\\
			Network & Number of Blocks & 1-4\\
			Block & Stride of Blocks & 1,2,4,8,16\\
			Block & Type &  residual, forward\\
			Block & Number of Convs & 1-4\\ 
			Layer & Kernel Size &  1,3,5,7,9,11 \\
			Layer & Output Channels & 4,8,..,64\\
			Layer & Activation Function & ReLU, None \\
			Accelerator & Array Size & $2\times2$, $4\times4$, $8\times8$, $16\times16$\\
            Accelerator & Memory Sizes & imputed \\
            Accelerator & Word Widths & imputed \\
            \hline 
		\end{tabular}
	\end{threeparttable}
\end{table}

The search for a target neural network and accelerator architecture configuration is
implemented as an evolution-based multi-objective optimization. As shown in Algorithm~\ref{alg:search},
the search first samples random neural network and hardware configuration from 
a joint search space $S$. The neural network is then trained using quantization aware training,
and the trained neural network is then evaluated on the validation set to obtain the accuracy metric. 
All further performance metrics are estimated by the analytical hardware model as described in Section~\ref{sec:deployment}. 
At the end of an optimization step, the sampled architecture parameters and the performance metrics are
added to the search history. After the initial population size $s$ has been reached, new architectures 
are derived from the current population using element-wise mutations. During search, HANNAH randomly samples 
from the following set of mutations:

\begin{enumerate}
\item Add/remove a block
\item Change block type between residual and feed-forward
\item Add/remove a convolutional layer 
\item Increase/decrease convolution size
\item Increase/decrease major block stride
\item Increase/decrease quantization word width
\item Increase/decrease MAC array size
\item Increase/decrease number of output channels
\end{enumerate}

To select the ancestor of the next evaluation point. Similar to current neural network 
search for TinyML on microcontrollers \cite{fedorov2019sparse,liberis2021munas}, we adopt 
randomized scalarization \cite{paria2020flexible}. 

In this approach the architecture parameters are ranked according to the following scalarization function: 
\begin{equation}
    f(M)\ =\ \max_{m_i\ \in\ M}\lambda_i\ \cdot\ m_i
\end{equation}
The $\lambda_i$ are sampled from the uniform distribution over $[0, \frac{1}{b_i}]$
where $b_i$ denotes a soft boundary for each target metric. Choosing $\lambda_i$ in 
this way ensures that candidates violating soft targets are always ranked behind 
targets that satisfy a target, while on the other hand encouraging the search to 
explore different parts of the search space near the Pareto boundary. 
\begin{algorithm}[t]
    \DontPrintSemicolon
    
      \KwInput{Design Space $S$, 
               Search Budget $b$, 
               Population Size $s$,
               Bounds $B$, 
               Training Set $T$,
               Validation Set $V$}
      \KwOutput{Search History $H$}
    
      \tcc{Start with empty search history}
      $H \leftarrow \emptyset$\\
      \For{$n \in \{1\ ...\ b\}$}
      {
        \If{$b\ \le\ s$}{
            \tcc{Sample Random Architecture and Accelerator}
            $a$ $\leftarrow$ sample\_random($S$)
        }
        \Else{
            \tcc{Sample Fitness Function from target bounds}
            $f$ $\leftarrow$  sample\_fitness\_function($B$)\\ 
            \tcc{Sort last $p$ candidates according to Fitness}
            $P$ $\leftarrow$ sort($H[-s:]$, $f$) \\
            \tcc{Apply random mutation to current best architecture}
            $a$ $\leftarrow$ mutate\_random(last(P)) 
          
        }
        \tcc{Quantization Aware Training}
        N $\leftarrow$ train($a$, $T$)\\
        \tcc{Evaluate trained architecture}
        error\_rate $\leftarrow$ evaluate($N$, $V$)\\
        \tcc{Estimate other metrics from hardware model}
        power, latency, area $\leftarrow$ hardware\_model($a$)
        \tcc{Append architecture and metrics to history}
        $H$.append($a$, (error\_rate, power, latency, area)) 
      }
    \caption{Neural architecture and Hardware Accelerator Search}
    \label{alg:search}
\end{algorithm}

\section{Experimental Evaluation}\label{sec:results}
The HANNAH framework has been implemented using current best practice libraries, using PyTorch version 1.10.1 \cite{paszke_pytorch_2019}.
  
Training hyperparameters have been set to fixed values for all experiments. All training runs use 30 epochs 
and use a batch size of 128 samples per minibatch. The optimizer used for training the network parameters 
is adamW with a one-cycle learning rate scheduling policy using a maximum learning rate of 0.005. 
All other optimizer and learning rate parameters are left at the default values provided by PyTorch.
All searches are run on a machine learning cluster using 4 Nvidia RTX2080 Ti GPUs. We train 8 neural network candidates in parallel which takes approximately 10 minutes. 
The training uses noisy quantization aware training\cite{fan2020training}.  The training sets are 
augmented using the provided background noise files for keyword spotting and voice activity detection and using random white noise for wake word detection. 
For inference the batch norm weights are folded into the weights and bias of the preceding convolutional layer \cite{jacob_quantization_2018}. 

As the hardware accelerator is set to operate at \SI{250}{\kilo\hertz} and the neural networks are trained and evaluated 
with an input time shift of \SI{100}{\milli\second} we set the latency constraint 
of the neural network architectures to \SI{25,000}{cycles} corresponding to the maximum input 
shift used during training and evaluation. The area 
constraints are set to \SI{150000}{\square\micro\metre} which is slightly less 
than the configuration used in \cite{bernardo2020ultratrail} for KWS.
The constraints for KWS are set to \SI{5}{\micro\watt} maximum power and accuracy constraints are set to \SI{93.0}{\%}. The other tasks use a power budget of  \SI{1}{\micro\watt} and accuracy constraints of \SI{95.0}{\%}.
All searches ran with a search budget of 3000 individual architectures and use a population size of 100.

We use the 22FDX technology by GlobalFoundries for implementation with low-leakage standard cells and SRAMs from Invecas. For synthesis and power estimation we use Cadence Genus 20.10 and Cadence Joules 21.11, respectively. For the average power consumption, we use two separate power estimations for the inference with previous feature loading and idle time. A weighted sum adds these two parts accordingly. The load of weights and bias is not included as it must be performed only once. It is evaluated at a \SI{25}{\degreeCelsius} TT corner with \SI{0.8}{\volt} supply voltage and no body bias voltage. The NPU uses clock-gating and low-power modes of the SRAM during idle times and waiting for the next inference to start.

\subsection{Results of Neural Network Co-Optimization}
\begin{figure}
	\centering
\includegraphics[width=0.9\columnwidth]{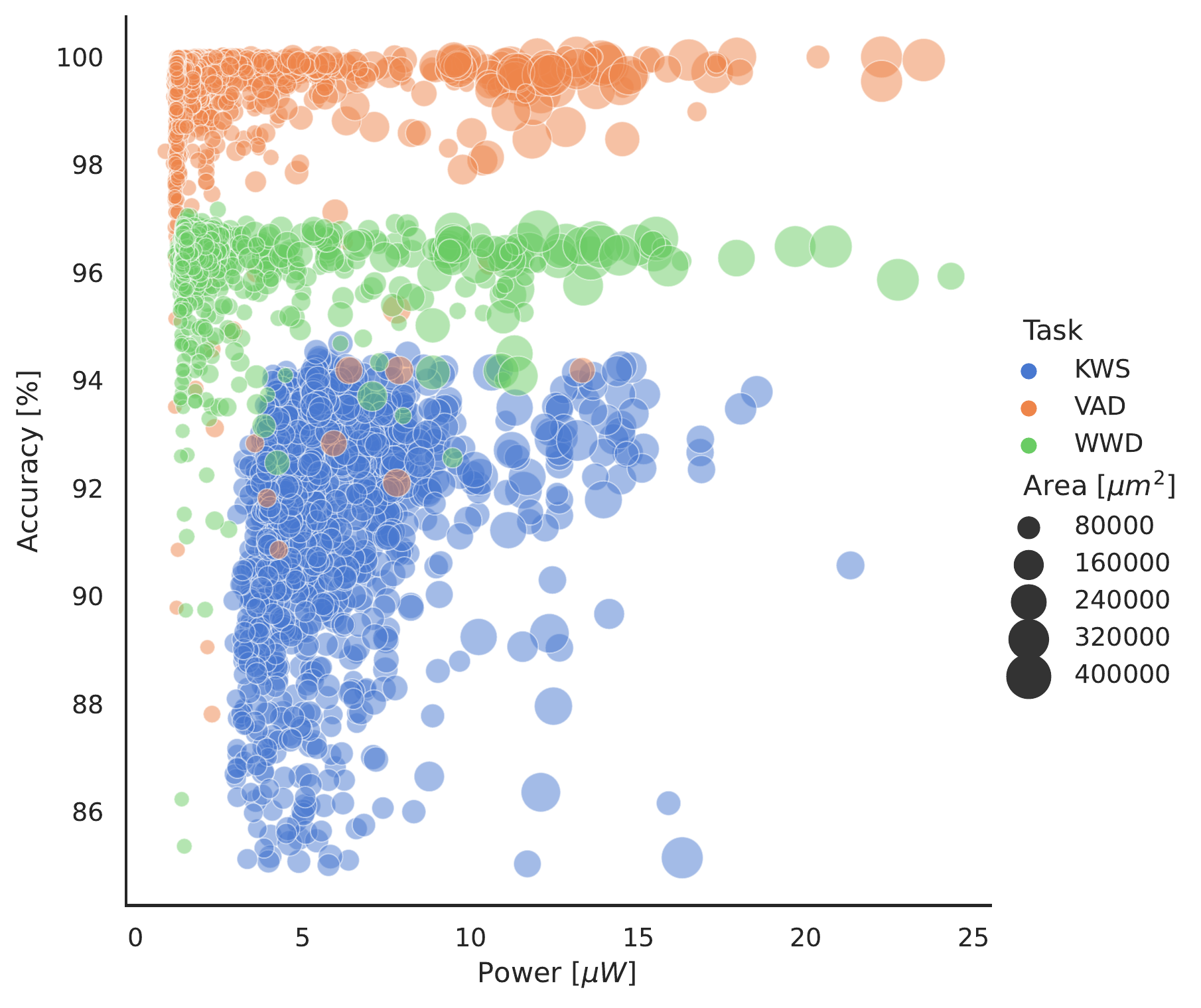}
\caption{Summarized power, area, and accuracy results}
\label{fig:nas_results}
\end{figure}

\begin{figure*}
	\centering
	\includegraphics[width=0.8\textwidth]{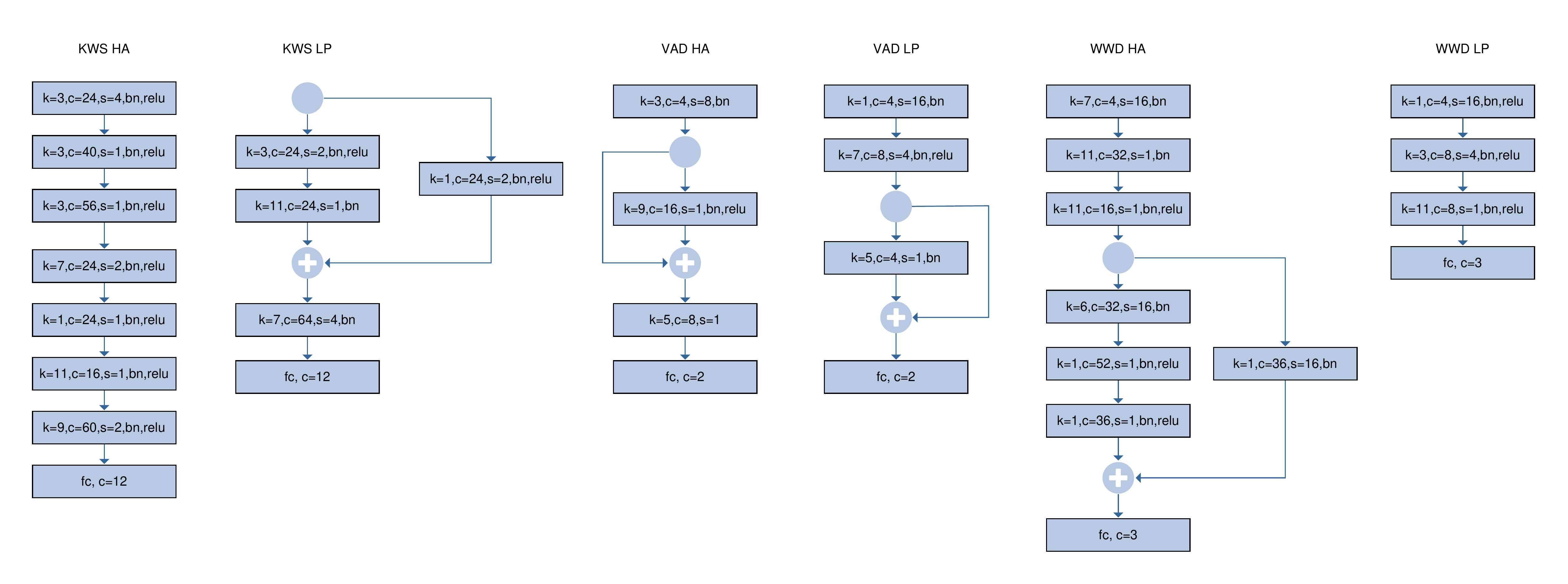}
	\caption{Low power and high accuracy network configurations for the three search tasks (fc = fully connected layer, bn = batch norm, relu = rectified linear unit, k=kernel size, c=number of output channels, s = stride)}
	\label{fig:results_search}
\end{figure*}

Figure~\ref{fig:nas_results} shows a summary of the NAS
results. For the three evaluation datasets, the search 
focuses the evaluated architectures around the Pareto front of power and accuracy,
and is able to find good trade-offs of power and accuracy.
For further analysis we select the Pareto points with lowest power (LP)  and
highest accuracy (HA) which are shown in Figure~\ref{fig:results_search}.
Half of the chosen networks are simple feed-forward convolutional neural networks, 
while the other half of neural networks contain at least one residual block. The
chosen networks sizes and operator configurations reflect the difficulty of the different 
machine learning tasks. The harder tasks use deeper networks, fewer strides and generally 
smaller convolution kernels, while the easier tasks use very shallow networks with rather
bit convolution kernels. These bigger filters are needed to compensate for the extremely big strides
used in the operator configurations. In all of the networks the search algorithm identified, possibilities
to introduce bottlenecks, e.g. convolutions with a larger number of output channels followed convolutions 
with comparably fewer output channels.   

\subsection{NPU Models}

\begin{figure}
	\centering
\includegraphics[width=0.9\columnwidth]{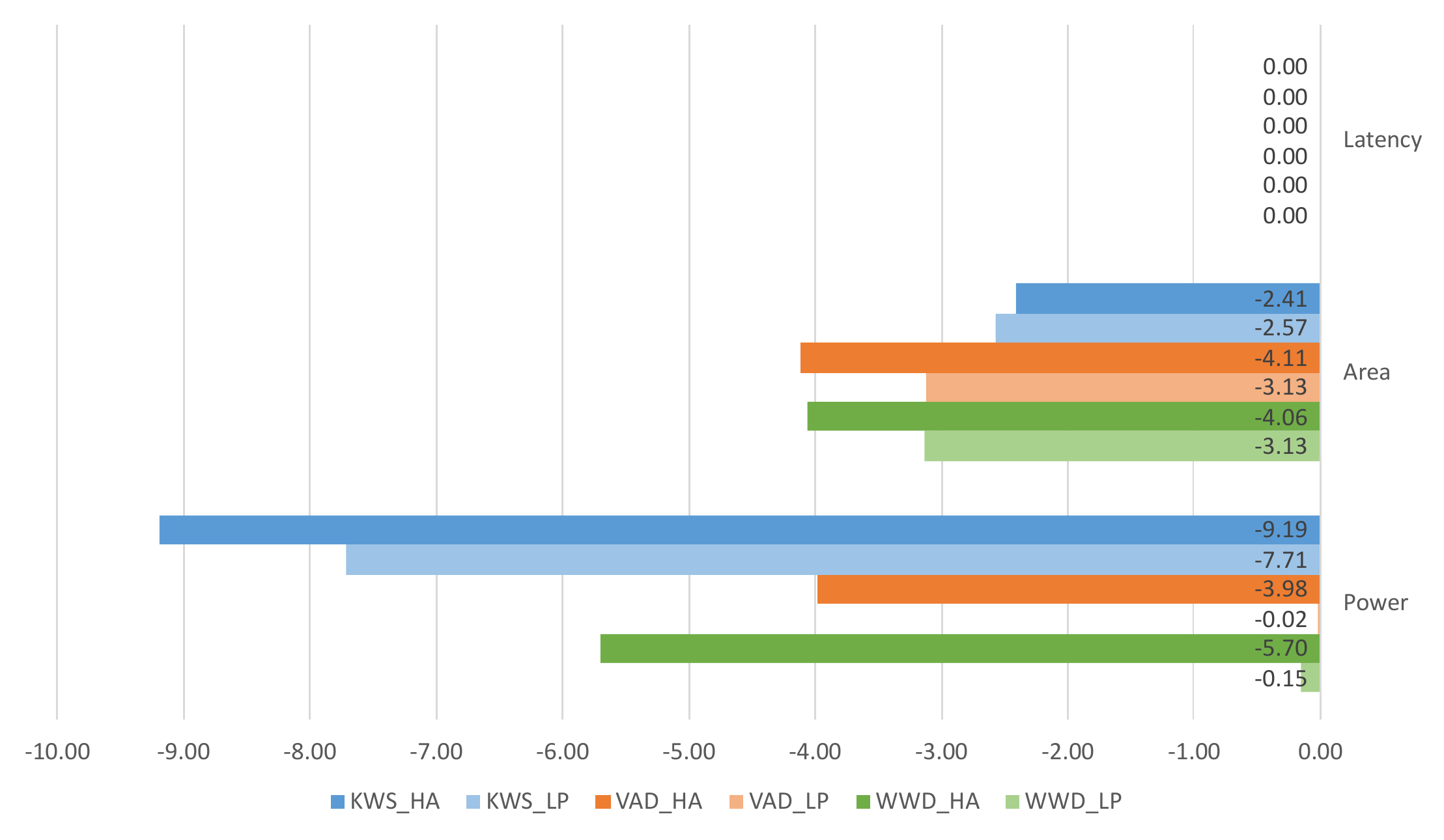}
\caption{Deviation of analytical models from gate-level simulation results in percentage points.}
\label{fig:comparison_models}
\end{figure}

Figure \ref{fig:comparison_models} shows the deviation in percentage points of the predictions of the analytical models and the data obtained from RTL simulation, synthesis and gate-level power estimation for latency, cell area and power consumption, respectively. 

The analytical performance model has no deviation as the behavior of the hardware accelerator is modelled exactly.

The area model slightly underestimates by $2.41$-$4.06$ percentage points because only the memory macros are exactly modelled and the other components do not handle quadratic growth. 
The deviation to the total synthesis area is additionally about \SI{3.6}{\percent}.

The power model uses the performance model to calculate the inference and idle time. Regarding the gate-level power consumption the power model results are in the range of $-9.19$-$-0.2$ percent points. Like the area model, the power model does not handle any non-linearity despite the array size. In addition, the chosen minimum value for the MAC array and OPU is conservative and has higher deviations for larger configurations like the ones for KWS.
Also, the model does not consider any data dependent or parasitic induced power consumption.
The NPU models are valid to guide the neural network co-optimization regarding performance and area with high precision.

\subsection{Comparison to related work}
Note,  that  due  to  the  differences  in datasets,  technology,  and  the  components  contained  in  the system, a direct comparison is only possible to a limited extent. All our implementations have a positive slack that allows at least a clock frequency of \SI{13.3}{\mega\hertz} without changing the voltage nor the synthesized netlist. This allows to reduce the latency but the power consumption would increase linearly with the clock frequency. For our work, the tables list the latency for feature loading and subsequent inference, as well as the latency with applied power gating. The corresponding power consumptions are also listed.

\begin{table*}[ht!]
	\centering
	\begin{threeparttable}
		\renewcommand{\arraystretch}{1.2}
		\caption{Comparison of Results on Keyword Spotting Task}
		\label{tab:results_kws}
		\begin{tabular}{c|c|c|c|c|c|c}
			\hline
			& \bfseries ESSCIRC'2018 \cite{giraldo2018laika} &\bfseries ISSCC'2020 \cite{giraldo_vocell_2020}  & \bfseries IEEE Access \cite{liu_ultra-low_2019} & \bfseries ESWEEK'2020 \cite{bernardo2020ultratrail}  & \bfseries High Accuracy & \bfseries Low Power \\
			\hline
			Technology 			 & \SI{65}{\nano\meter} & \SI{65}{\nano\meter} & \SI{22}{\nano\meter} & \SI{22}{\nano\meter} & \SI{22}{\nano\meter} & \SI{22}{\nano\meter} \\
 			Area 				 & \SI{1.03}{\square\milli\meter} \tnote{c} & \SI{2.56}{\square\milli\meter} \tnote{c} & \SI{0.75}{\square\milli\meter} \tnote{c} & \SI{0.20}{\square\milli\meter} \tnote{c} & \SI{0.126}{\square\milli\meter} \tnote{d} & \SI{0.118}{\square\milli\meter} \tnote{d}\\
			Frequency 			 & \SI{250}{\kilo\hertz} & \SI{250}{\kilo\hertz} & \SI{250}{\kilo\hertz} & \SI{250}{\kilo\hertz} & \SI{250}{\kilo\hertz} & \SI{250}{\kilo\hertz} \\
			Latency 			 & \SI{16}{\milli\second} & \SI{16}{\milli\second} & \SI{20}{\milli\second}  & \SI{100}{\milli\second} & \Shortunderstack{\SI{100}{\milli\second}\\\SI{35.5}{\milli\second}\tnote{e}} &\Shortunderstack{\SI{100}{\milli\second}\\\SI{19.0}{\milli\second}\tnote{e}}\\
			Voltage 			 & \SI{0.6}{\volt} & \SI{0.6}{\volt} & \SI{0.55}{\volt} & \SI{0.8}{\volt} & \SI{0.8}{\volt} & \SI{0.8}{\volt} \\
			DNN Structure 		 & LSTM+FC & LSTM+FC & CONV+FC & TC-ResNet & CONV+FC & TC-ResNet\\
			Word Width (Weights) & 4/8 & 4/8 & 7 & 6 & 6 & 6 \\
			Word Width (Inputs)  & 8   & 8   & 8 & 8 & 6 & 6 \\
            MAC Array Size       & -   & -   & - & $8\times8$ & $8\times8$ & $8\times8$ \\
			Accuracy\tnote{a}    & - & \SI{90.87}{\percent} (GSCD) & \SI{90.51}{\percent} (GSCD) & \SI{93.09}{\percent} (GSCD)  & \SI{94.73}{\percent} (GSCD) & \SI{93.54}{\percent} (GSCD)\\
			F1-Score\tnote{a}    & \SI{90.00}{\percent} (TIMIT)\tnote{b} & - & - &  - & \SI{94.73}{\percent} (GSCD) & \SI{93.54}{\percent} (GSCD)  \\
			Keywords 			 & 4 & 10 & 10 & 10  & 10 & 10 \\
			Power 				 & \SI{5.0}{\micro\watt} & \SI{10.6}{\micro\watt} & \SI{52}{\micro\watt} & \SI{8.2}{\micro\watt} & \Shortunderstack{\SI{4.73}{\micro\watt}\\\SI{10.4}{\micro\watt\tnote{f}}} & \Shortunderstack{\SI{4.17}{\micro\watt}\\\SI{10.7}{\micro\watt}\tnote{f}} \\
			\hline
		\end{tabular}
		\begin{tablenotes}
			\item [a] SNR $\geq$ \SI{1000}{\decibel}
			\item [b] Note the difference in datasets
            \item [c] Area for layout/chip
            \item [d] Cell area for synthesis
			\item [e] Active inference latency
			\item [f] Power when active
		\end{tablenotes}
	\end{threeparttable}
\end{table*}

\begin{table*}
	\centering
	\begin{threeparttable}
		\renewcommand{\arraystretch}{1.2}
		\caption{Comparison of Results on Voice Activity Detection Task}
		\label{tab:results_vad}
		\begin{tabular}{c|c|c|c|c|c}
			\hline
			& \bfseries ISSC'2017 \cite{price_low-power_2018} & \bfseries JSSC'2019\cite{yang2019design} & \bfseries GLSVLSI'2020\cite{liu2020background} & \bfseries High Accuracy & \bfseries Low Power \\ 
			\hline
			Technology & \SI{65}{\nano\meter} & \SI{28}{\nano\meter} & \SI{180}{\nano\meter} & \SI{22}{\nano\meter} & \SI{22}{\nano\meter}\\
 			Area & - & - & ~\SI{0.62}{\square\milli\meter} (estimated NPU) & \SI{0.050}{\square\milli\meter}  & \SI{0.036}{\square\milli\meter} \\
			Frequency & \SI{1.68}{\mega\hertz}-\SI{47.8}{\mega\hertz} & - & - & \SI{250}{\kilo\hertz} & \SI{250}{\kilo\hertz} \\
			Latency & \SI{100}{\milli\second}-\SI{500}{\milli\second} & \SI{10}{\milli\second} & \SI{10}{\milli\second} & \Shortunderstack{\SI{100}{\milli\second}\\\SI{9.1}{\milli\second}\tnote{a}} & \Shortunderstack{\SI{100}{\milli\second}\\\SI{4.5}{\milli\second}\tnote{a}}\\
			Voltage & \SI{0.5}{\volt}-\SI{0.9}{\volt} & \SI{0.55}{\volt} & \SI{0.55}{\volt} & \SI{0.8}{\volt} & \SI{0.8}{\volt}\\
			DNN Structure & FC & FC & FC & TC-ResNet & TC-ResNet \\
			Word Width (Weights) & 16 (from \cite{yang2019design}) & 1     & 1                         & 4 & 2\\
			Word Width (Inputs)  & 16 (from \cite{yang2019design}) & 1/4/8 & 9 (first layer), 1 (else) & 4 & 4 \\
            MAC Array Size     & - &- & - & $4\times4$ & $4\times4$ \\
			Accuracy & \Shortunderstack{\SI{10.0}{\percent} EER @\SI{7}{\decibel} \\ (AURORA2)} & \Shortunderstack{95/\SI{92.0}{\percent} @\SI{10}{\decibel} \\ 90/\SI{87.0}{\percent} @\SI{5}{\decibel} \\85/\SI{80.0}{\percent} @\SI{-5}{\decibel}\\(TIMIT, Noise-92)} & \Shortunderstack{84/\SI{85}{\percent} @\SI{10}{\decibel}\\(Aurora4, DEMAND) } &  \Shortunderstack{\SI{99.88}{\percent} \\ \SI{99.43}{\percent} @\SI{10}{\decibel} \\ \SI{99.40}{\percent} @\SI{5}{\decibel} \\ \SI{98.04}{\percent} @\SI{0}{\decibel}\\(UW/NU,TUT)} &  \Shortunderstack{\SI{97.17}{\percent} \\ \SI{95.37}{\percent} @\SI{10}{\decibel} \\ \SI{93.04}{\percent} @\SI{5}{\decibel} \\ \SI{88.90}{\percent} @\SI{0}{\decibel}\\(UW/NU,TUT)}\\
			Power & \SI{22.3}{\micro\watt} & 2/5/\SI{8}{\micro\watt} & \Shortunderstack{\SI{1.01}{\micro\watt}@AFE+VAD \\ \SI{0.63}{\micro\watt}@VAD only} &\Shortunderstack{\SI{851}{\nano\watt}\\\SI{4.17}{\micro\watt}\tnote{b}} & \Shortunderstack{\SI{435}{\nano\watt}\\\SI{2.13}{\micro\watt}\tnote{b}}\\
			\hline
		\end{tabular}
		\begin{tablenotes}
			\item [a] Active inference latency
			\item [b] Power when active
		\end{tablenotes}
	\end{threeparttable}
\end{table*}

\begin{table*}
	\centering
	\begin{threeparttable}
		\renewcommand{\arraystretch}{1.2}
		\caption{Comparison of Results on Wake Word Detection Task}
		\label{tab:results_wwd}
		\begin{tabular}{c|c|c|c|c}
			\hline
															& \bfseries arXiv'2020\cite{wang2020always} & \bfseries VLSI'2019\cite{guo20195} 							& \bfseries High Accuracy 												& \bfseries Low Power\\
			\hline					
			Technology                 						& \SI{65}{\nano\meter} 						& \SI{65}{\nano\meter} 											& \SI{22}{\nano\meter} 													& \SI{22}{\nano\meter}\\
 			Area              								& \SI{1.99}{\square\milli\meter}			& \SI{6.2}{\square\milli\meter} (FE+NPU) 						& \SI{0.046}{\square\milli\meter} 										& \SI{0.036}{\square\milli\meter}\\
			Frequency                  					 	& \SI{70}{\kilo\hertz} 						& 5-\SI{75}{\mega\hertz}										& \SI{250}{\kilo\hertz} 												& \SI{250}{\kilo\hertz} \\
			Latency                    					 	& - 										& \SI{3.36}{\micro\second}-\SI{1.91}{\milli\second}				& \Shortunderstack{\SI{100}{\milli\second}\\\SI{18.2}{\milli\second}\tnote{a}} 	& \Shortunderstack{\SI{100}{\milli\second}\\\SI{5.4}{\milli\second}\tnote{a}}\\
			Voltage                    					 	& 0.5-\SI{1.0}{\volt}						& 0.9-\SI{1.1}{\volt} 											& \SI{0.8}{\volt} 														& \SI{0.8}{\volt}\\
			DNN Structure              					 	& SNN 										& RNN 															& TC-ResNet																& CONV+FC\\
			\Shortunderstack{Word Width\\(Weights)}        	& 6											& 1 															& 4 																	& 2\\
			\Shortunderstack{Word Width\\(Inputs)}         	& 1 										& 1 															& 4 																	& 4\\
            MAC Array Size             						& -											& - 															& $4\times4$ 															& $4\times4$\\
			Accuracy 										& \Shortunderstack{\SI{95.8}{\percent}@\SI{40}{\decibel} \\ \raisebox{-0.75ex}{\~{}}\SI{90.5}{\percent}@\SI{0}{\decibel}}& \SI{91.9}{\percent} 										& \Shortunderstack{\SI{96.81}{\percent} \\ \SI{95.8}{\percent}@\SI{40}{\decibel} \\ \SI{94.6}{\percent}@\SI{0}{\decibel}} & \Shortunderstack{\SI{95.64}{\percent} \\ \SI{95.3}{\percent}@\SI{40}{\decibel} \\ \SI{95.1}{\percent}@\SI{0}{\decibel}}\\
			Power                      						&  75-\SI{220}{\nano\watt} 					& \SI{134}{\micro\watt} 			& \Shortunderstack{\SI{998}{\nano\watt}\\\SI{4.15}{\micro\watt}\tnote{b}} 				& \Shortunderstack{\SI{435}{\nano\watt}\\\SI{2.14}{\micro\watt}\tnote{b}} \\
			\hline
		\end{tabular}
		\begin{tablenotes}
			\item [a] Active inference latency 
			\item [b] Power when active
		\end{tablenotes}
	\end{threeparttable}
\end{table*}

\subsubsection{Keyword Spotting}
Table \ref{tab:results_kws} lists state-of-the-art KWS accelerators and systems. \cite{giraldo2018laika} uses a recurrent neural network with LSTM units to detect only four keywords and has a higher power consumption than both of our variants. \cite{giraldo_vocell_2020} proposes a system with feature extraction and a hierarchical accelerator setup including a sound detector, a KWS and a speaker validation module to reduce power consumption. The system in \cite{liu_ultra-low_2019} also includes a feature extraction but nevertheless with \SI{41.3}{\micro\watt} has a higher power consumption than all other works. Our LP and HA variants are based on \cite{bernardo2020ultratrail} and achieve the highest accuracy on GSCD while both variants still have a lower power consumption.

\subsubsection{Voice Activity Detection}
State-of-the-art VAD accelerators and systems are shown in Table \ref{tab:results_vad}. Note that almost all chosen datasets, technologies and latencies are different.
Price et al. \cite{price_low-power_2018} uses a VAD accelerator for a fully connected neural network as a preliminary stage with a power consumption of \SI{22.3}{\micro\watt} and a latency of at least \SI{100}{\milli\second}.
\cite{yang2019design} presents an ultra-low power VAD accelerator for binary neural networks. The power consumption varies between $2$ and \SI{8}{\micro\watt} depending on the used word width with speech accuracy of up to \SI{95}{\percent} with \SI{10}{\decibel} of restaurant noise.
\cite{liu2020background} uses an analogue front end (AFE) to calculate the features and so the overall system consumes just about \SI{1}{\micro\watt}. The accelerator itself has a low latency and power consumption but also a low accuracy compared to other works. Our variants have the lowest power consumption and highest accuracies without noise and a latency of \SI{100}{\milli\second}.

\subsubsection{Wake Word Detection}
State-of-the-art WWD accelerators and systems are shown in Table \ref{tab:results_wwd} using the ``Hey Snips" dataset. \cite{wang2020always} uses spiking neural networks combined with event-driven clock and power gating to reduce the power consumption. This is the only state of the art accelerator that can beat our work in some of the accuracy and power values, but needs a very specialized neural network and hardware accelerator architecture. \cite{guo20195} proposes a system with accelerators for feature extraction, VAD and binary RNNs. This system has a lower accuracy and higher power consumption than our accelerators. 

\section{Conclusion and Future Work}\label{sec:conclusion}

In this paper, we have presented HANNAH, a framework for co-optimization of neural networks and 
corresponding hardware accelerators. The framework is able to find neural networks to 
automatically find and generate ultra-low power implementations on a set of three speech 
processing benchmarks. In our future work we intend to extend this work to support heterogeneous
deployments and a larger class of DNNs for intelligent sensor processing and perception in edge devices. 


\bibliographystyle{IEEEtran}
\bibliography{references}
\end{document}